\documentstyle[12pt, epsf, rotate, amssymb]{article}
\begin{document}

\begin{center}
\begin{Large}
{\bf Restrictions on parameters of power-law magnetic field decay for
accreting isolated neutron stars}


S.B. Popov, M.E. Prokhorov

\end{Large}

Sternberg Astronomical Institute, Moscow, Russia

119899, Universitetskii pr. 13

polar@xray.sai.msu.su, mystery@sai.msu.su

\vskip 0.1cm

{\bf Abstract}

 \end{center}

 In this short note we discuss the influence of power-law magnetic field
decay on the evolution of old accreting isolated neutron stars.
We show, that, contrary to exponential field decay (Popov \& Prokhorov
2000), no additional restrictions  
can be made for the parameters of power-law decay 
from the statistics of isolated neutron star candidates in ROSAT
observations.

We also briefly discuss the fate of old magnetars with and without field
decay, and describe parameters of old accreting magnetars.  


\section{Introduction}

Isolated neutron stars (INSs), which don't show radio pulsar activity 
attract now much attention of astrophysicists
due to recent observations of several candidates with the ROSAT sattelite
(see Ne\"uhauser \& Tr\"umper 1999 and
a review in Treves et al. 2000). As we discussed in our previous paper
(Popov \& Prokhorov 2000) INSs can be important for discussion of different
models of magnetic field decay (MFD) in NSs in general.

During its evolution an INS can pass through four phases: {\it ``ejector'',
``propeller'', ``accretor''} and {\it ``georotator''}. 
At the first stage the INS is spinning down
according to the magneto-dipole formula till  so-called ejector period is
reached. At the second stage captured matter cannot penetrate down to the
surface of the INS, and the star continue to spin down faster than
at the stage of ejection. At last, so-called accretor period is reached,
and matter can fall down: accretion starts. If the INS's velocity (or
magnetic field) is high enough, the star can appear as a georotator, where
matter cannot be captured, as far as the magnetosphere
radius is large than the radius of gravitational capture.

Several models of MFD in NSs were suggested during the last
20-30 years
(see for example a recent brief review by Konar \& Bhattacharya).
Most of these models can be fitted by exponential or power-law decay,
or by their combination with some set of parameters. 
INSs can be an important class of
objects for verification of different theories of MFD, because in
these sources accretion rate is negligible, so it is not necessary to take
into account the influence of accretion onto MFD (Urpin et al. 1996). 
Spin-up/spin-down rates on the stage of accretion 
are also relatively low in comparison with NSs in binary systems. It means,
that in INSs MFD operates in the ``purest'' form (Popov \&
Konenkov 1998). That's why
these objects have a special importance, in our opinion, for investigations
of observational appearance of different effects of MFD. 

Recently, Colpi et al. (2000) discussed power-law models of MFD 
in INSs and applied them to highly magnetized
NSs, {\it ``magnetars''}. Here we briefly discuss later stages of evolution
of INSs with the power-law MFD, and estimate if it is possible for them to
reach the stage of accretion, and if yes, what can be their properties at
this stage.

Our analysis follows the papers Popov \& Prokhorov (2000), 
and Colpi et al. (2000). So, we just repeat calculations of Popov \&
Prokhorov (2000) but for the power-law decay, using some results of Colpi et al.
(2000). And we refer to these papers for all details of terminology,
calculations etc.

\section{Power-law decay}

Power-law (as also exponential) MFD is a widely discussed variant of
NSs' field evolution. Power-law is a good fit for several different
calculations of the field evolution 
(Goldreich \& Reisenegger 1992, Geppert et al. 2000).
The power-law MFD can be described with the following 
simple formula (Colpi et al. 2000):

\begin{equation}
\frac{dB}{dt}=-aB^{1+\alpha}.
\end{equation}
So, we have only two parameters of decay: $a$ and $\alpha$. 
As far as this decay is relatively slow for the most interesting
values of $\alpha
\gtrsim 1$ (we use the same units as in Colpi et al. 2000),
we don't specify any bottom magnetic field, contrary to what we made for
more rapid exponential decay (Popov \& Prokhorov 2000). Even for the Model C
from Colpi et al. (2000) (see Table 1) 
with relatively fast MFD the magnetic field can decrease only
down to $\sim 10^8$ G in $10^{10}$ yrs (see Fig. 1).
But for $\alpha <1$ the magnetic 
field can decay significantly during the Hubble time (we call
here {\it ``the Hubble time''} time interval $10^{10}$ yrs, which is nearly equal to
the age of our Galaxy) for any reasonable value of $a$. And, probably, it is
useful to introduce in the later case a bottom field.

At the stage of ejection an INS 
is spinning down according to the magneto-dipole formula: \\
$P\dot P \approx b B^2$.
Here (and everywhere below) $b=3$, values of
magnetic field, $B$, $B_{\infty}$ and $B_0$, 
are taken in units $10^{13}$ G and time, $t$, in units $10^6$ yrs (as in Colpi
et al. 2000).

In the table we show parameters of the Models A, B, C from Colpi et al. (2000).
$B_{\infty}$ is the  magnetic field calculated for $t=t_{Hubble}=10^{10}$ yrs and
for the initial field $B_0=10^{12}$ G. Models A and B correspond to
ambipolar diffusion in the irrotational and the solenoidal modes respectively. 
Model C describes MFD in the case of the Hall cascade.

\begin{table}[h]
\caption[]{Models A,B,C from Colpi et al. (2000)}
\centerline{
\begin{tabular}{|c||c|c|c|}
\hline
Model & A & B & C\\
\hline
\hline
$a$ & 0.01 & 0.15 & 10\\
$\alpha$ & 5/4 & 5/4 & 1 \\
$B_{\infty}$ & $\approx 1.9 \cdot 10^{11}$ G & $\approx 2.4 \cdot 10^{10}$ G
& $\approx 10^8$ G \\
\hline
\end{tabular}
}
\end{table}

In Fig. 2 we show dependence of the ejector period, $p_e$, and the
asymptotic period, $p_{\infty}$, on the parameter $a$ for $\alpha=1$ 
for different values of the initial magnetic field, $B_0$:

\begin{equation}
p_e=25.7\, B_{\infty}^{1/2}n^{-1/4}v_{10}^{1/2} \, {\rm s},
\end{equation}

\begin{equation}
p_{\infty}^2=\frac{2}{2-\alpha}\frac{b}{a}B_0^{2-\alpha}.
\end{equation}
Here $v_{10}$ is velocity $(v_{INS}^2+v_s^2)^{1/2}$ in units 10 km/s;
$v_{INS}$ is the spatial velocity of the INS and $v_s$ - sound velocity.
$n$ is the interstellar medium (ISM) number density. $B_0$ - initial magnetic
field.\\
$p_e$ was calculated for $t=t_{Hubble}=10^{10}$ yrs,
i.e. for the moment, when $B=B_{\infty}$.

\begin{figure}
\epsfxsize=0.9\hsize
\centerline{\rotate[r]{\epsfbox{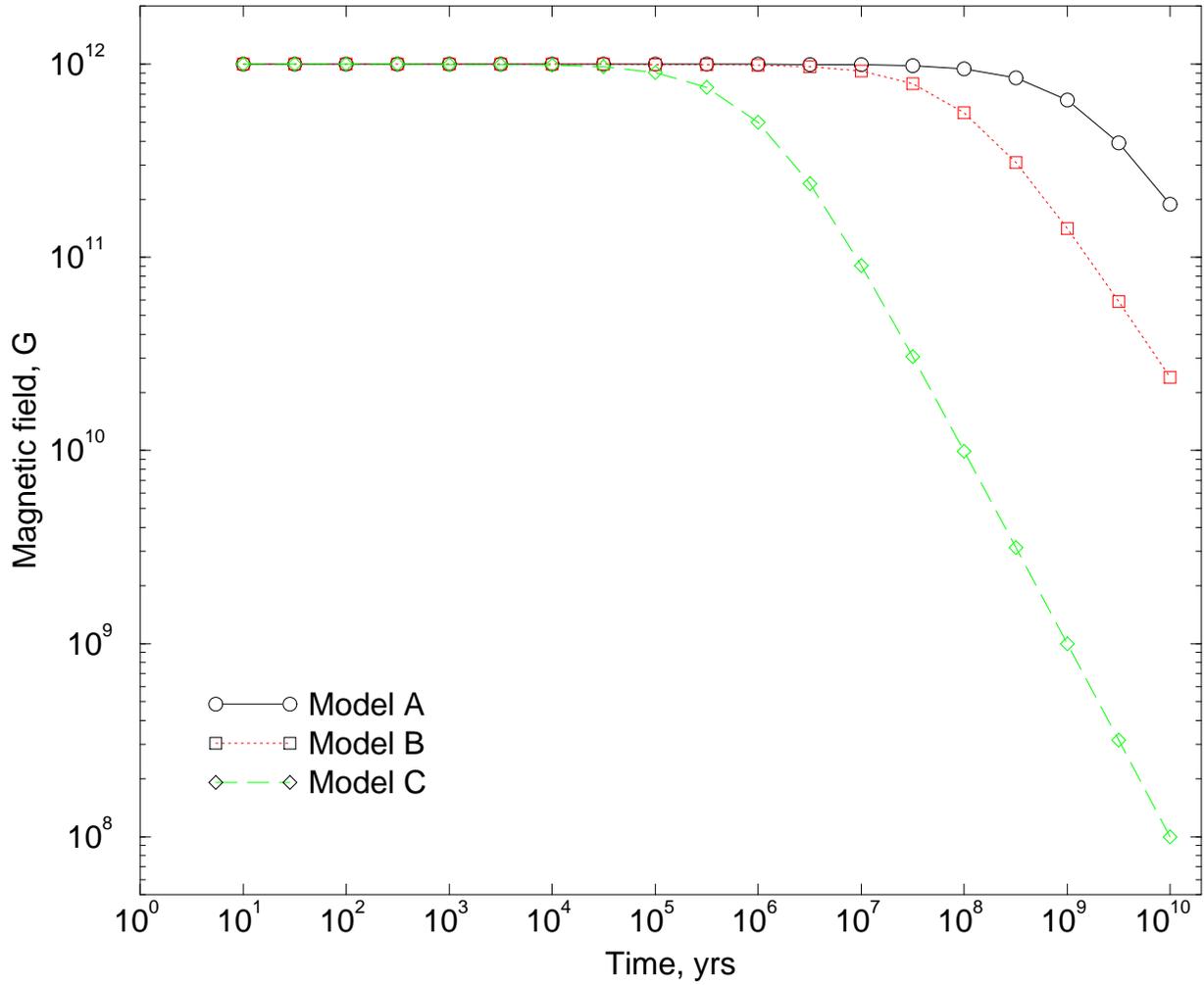}}}
\caption{Power-law MFD. Model A: $a=0.01, \alpha=1.25$; solid line
with circles.
Model B: $a=0.15, \alpha=1.25$; dashed line with squares.
Model C: $a=10,\, \alpha=1$; long-dashed line with diamonds.
Models were described in details in Colpi et al. (2000) (see also Table 1).}  
\end{figure}

\begin{figure}
\epsfxsize=0.9\hsize
\centerline{\rotate[r]{\epsfbox{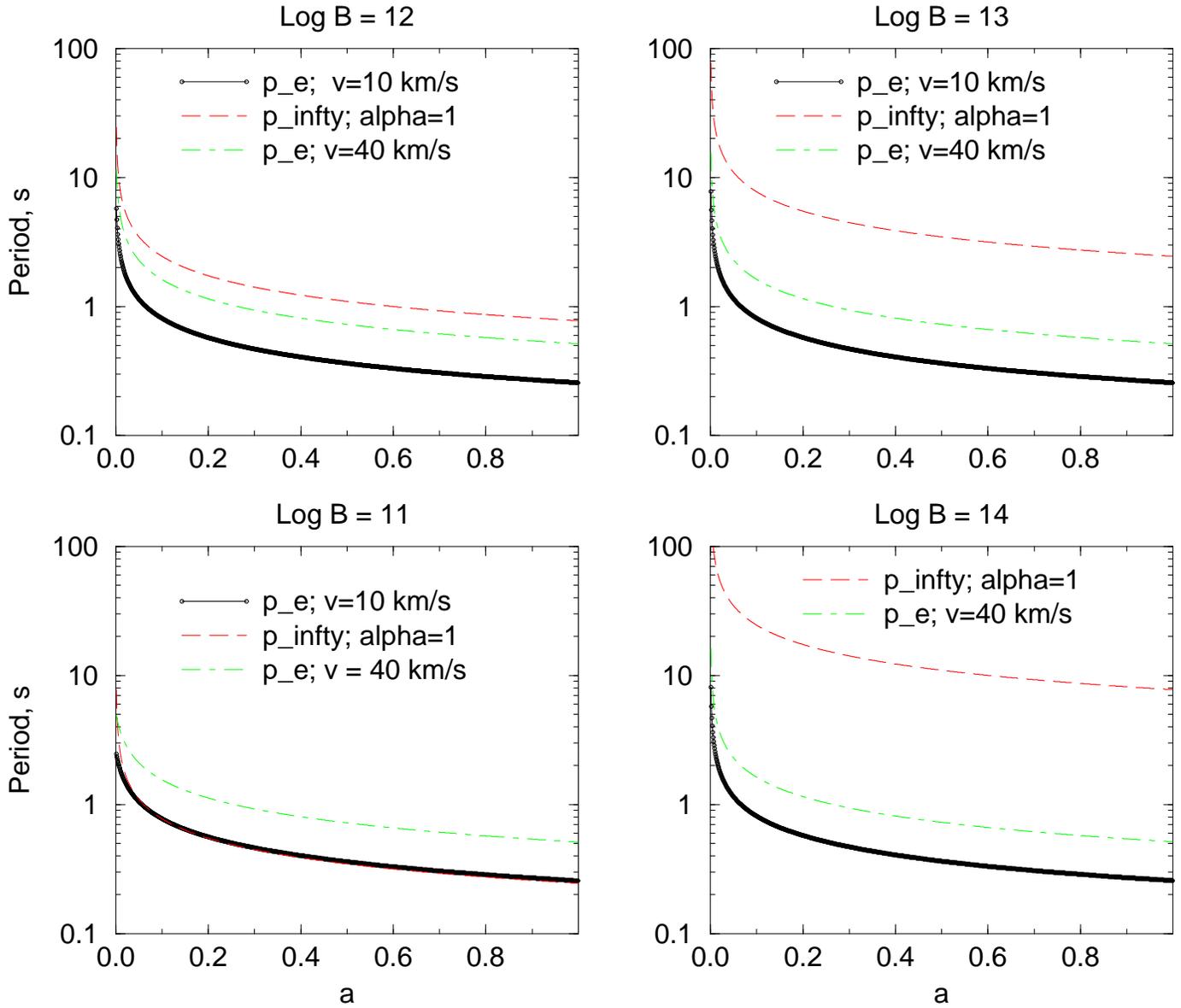}}}
\caption{Periods vs. parameter $a$ for different values of the initial magnetic
field: $10^{11}, 10^{12}, 10^{13}, 10^{14}$~G.}
\end{figure}

\begin{figure}
\epsfxsize=0.9\hsize
\centerline{\rotate[r]{\epsfbox{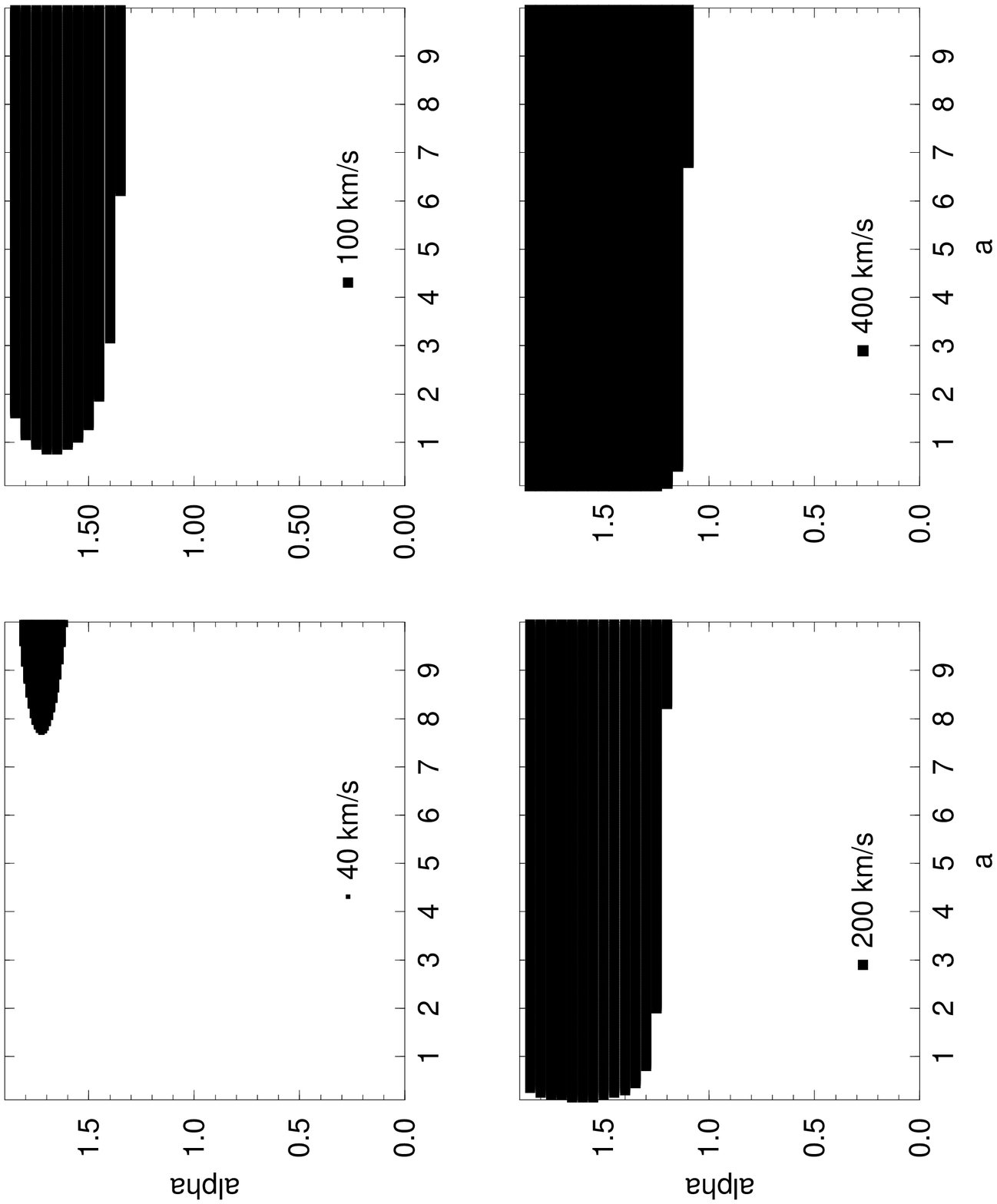}}}
\caption{``Forbidden'' regions for the initial field $10^{13}$ G and different
INS's spatial velocities: 40 km/s, 100 km/s, 200 km/s and 400 km/s. 
In the filled regions NSs never leave the ejector stage.}  
\end{figure}

It is clear from  Fig. 2, 
that for the initial field  $\gtrsim 10^{11}$ G low velocity INSs
are able to come to the stage of accretion: for $B_0=10^{11}$ G lines for 
$p_{\infty}$ and $p_e$ for the lowest possible velocity, 10 km/s, coincides.

In  Fig. 3 we show ``forbidden'' regions on the plane $a$--$\alpha$,
where an INS for a given velocity for sure
cannot come to the stage of accretion in the
Hubble time (compare with ``forbidden'' regions in Popov \& Prokhorov 2000). 
In a forbidden region an INS for specified parameters cannot leave the stage of
ejector even after $10^{10}$ years of evolution. 
If one also takes into account the stage
of propeller (between ejector and accretor stages) it becomes clear, that 
``forbidden''' regions for an INS which cannot reach the stage of accretion is
even larger. We note, that the propeller stage can be shorter 
(probably much shorter, especially for constant field) than
the stage of ejection (see Lipunov \& Popov 1995 for detailed arguments),
so the ``forbidden'' regions on Fig.~3 cannot become much larger if one also
takes into account the stage of propeller. It is also important, that we
take very low INS's velocity and high ISM density. For most part of INSs all
plotted ``forbidden'' regions should be larger.

One can see, that for the most interesting cases (Models A, B, C
from Colpi et al. 2000) and $v<200$ km/s INSs can reach the stage of accretion.
It is an important point, that fraction of low velocity NSs is very
small (Popov et al. 2000) and most of them have velocities about 200 km/s.

\section{Evolved magnetars}

In the last several years a new class of objects - highly magnetized NSs,
{\it ``magnetars''} (Duncan \& Thompson 1992) -- became very popular in
connection with soft $\gamma$-repeaters (SGR) and anomalous X-ray pulsars
(AXP) (see Mereghetti \& Stella 1995,   
Kouveliotou et al. 1999, Mereghetti 1999 and recent theoretical works Alpar 1999, 
Marsden et al. 2000, Perna et al. 2000).

 Magnetars come to the propeller stage with periods $\sim 10$ -- $ 100$ s in
the Models A, B, C (see Fig. 2 in Colpi et al. 2000).
Then their periods quickly increase, and NSs come to the stage of accretion with
significantly longer periods, and at that stage they evolve to a so-called
equilibrium period (Lipunov \& Popov 1995, 
Konenkov \& Popov 1997) due to accretion of the turbulent ISM:
\begin{equation}
p_{eq}\sim 2800 B^{2/3}I_{45}^{1/3}n^{-2/3}v_{10}^{13/3}
v_{t_{10}}^{-2/3}M_{1.4}^{-8/3}\, {\rm s}
\end{equation}
Here $v_t$ is a characteristic turbulent velocity, $I$ -- moment of inertia,
$M$ -- INS's mass.

Isolated accretor can be observed both with positive and negative sign of
$\dot p$ (Lipunov \& Popov 1995). Spin periods of INSs can differ significantly
from $p_{eq}$
contrary to NSs in disc-fed binaries, and similar to NSs in wide binaries,
where accreted matter is captured from giant's stellar wind. It happens
because spin-up/spin-down moments are relatively small.

As the field is decaying the equilibrium period is decreasing, coming to
28 sec when the field is equal to $10^{10}$ G (we note here recently discovered
objects RX J0420.0-5022 (Haberl et al. 2000) with spin period $\sim 22.7$ s).

It is important to discuss the possibility, that evolved magnetar can appear
as georotator (see Lipunov 1992 for detailed description or Popov et al. 2000 
for short description of different INSs' stages). 
It happens if:

\begin{equation}
v\gtrsim 300 B^{-1/5} n^{1/10} \, {\rm km/s}.
\end{equation}

For all values of $a$ and $\alpha$ that we used (see Fig. 3) 
NSs, at the end of their evolution ($t=10^{10}$ yrs), 
have magnetic fields $\lesssim 10^{12}$ G for wide range of initial
fields, so they never appear as
georotators if $v<580$ km/s for $n=1 {\rm cm}^{-3}$.
But without MFD magnetars with $B\gtrsim 10^{15}$ G and velocities 
$v\gtrsim 100$~km/s can appear as georotators. 

In Popov et al. (2000) it was shown, that georotator is a rare stage for
INSs, because an INS can come to the georotator stage only from the propeller
or accretor stage, but all these phases require relatively low velocities, and
high velocity INSs spend most of their lives as ejectors. 
This situation is opposite to binary systems, where a lot of
georotators are expected for fast stellar winds (wind velocity can be much
faster than INS's velocity relative to ISM).

Without MFD magnetars also can appear as accreting sources.
In that case they can have very long periods and very narrow accretion
columns (that means high temperature). 
Such sources are not observed now. Absence of some
specific sources associated with evolved magnetars (binary or isolated)
can put some limits on
their number and properties (dr. V. Gvaramadze drew our attention to this
point). 

At the accretion part of INSs' evolution 
periods stay relatively close to $p_{eq}$ (but can fluctuate around this
value), and INSs' magnetic fields decay down to
$\sim 10^{10}-10^{11}$ G in several billion years for the Models A and B. 
It corresponds to the polar cap radius
about 0.15 km and temperature about 250 -- 260 eV, higher than for the observed
INS candidates with temperature about 50 -- 80~eV. We calculate the polar
cap radius, $R_{cap}= R\sqrt{(R/R_A)}$, with the following formula: 
\begin{equation}
R_{cap}=
6\cdot 10^3\, B^{-2/7}n^{1/7} v_{10}^{-3/7}\ {\rm cm}.   
\end{equation}
Here $R_A\simeq 1.8\cdot 10^{10}  n^{-2/7}
v_{10}^{6/7}B^{4/7}\, {\rm cm}$ is the Alfven radius.
The temperature can be even larger, than it follows from the formula above
as far as for very high field matter can be channeled in a narrow ring, so
the area of the emitting region will be just a fraction of the total polar
cap area.

As the field is decreasing the radius of the polar cap is increasing, and
the temperature is falling.
Sources with such properties (temperature about 250-260 eV)
are not observed yet (Schwope et al. 1999). But if the number of
magnetars is significant (about 10\% of all NSs) accreting evolved magnetars 
can be found in the near
future, as far as now we know about 5 accreting INS candidates (Treves et
al. 2000, Ne\"uhauser \& Trumper 1999),
and their number can be increased in future. $\dot p$ measurements are
necessary to  understand the nature of such sources, if they are observed.

 Recently discovered object RX J0420.0-5022 (Haberl et al. 2000) with the spin
period $\sim 22.7$~s, can be an example of an INS with  decayed magnetic
field accreting from the ISM, as previously RX J0720.4-3125. 
Due to relatively low temperature, 57 eV, its
progenitor cannot be a magnetar for power-law MFD (Models A,B,C) or
similar sets of parameters, because a very large polar cap is needed, which
is difficult to obtain in these models. 
Of course RX J0420.0-5022 can be explained also as a cooling NS. 
The question "are the observed candidates cooling or accreting objects?" is
still open (see Treves et al. 2000). If one finds an
object with $p\gtrsim100$ s and temperature about 50 -- 70~eV  can be a strong
argument for its accretion nature, 
as far as such long periods for magnetars
can be reached only for very high initial magnetic fields (see Fig. 2 in
Colpi et al. 2000) for reasonable models of MFD and other parameters.

\section{Conclusions}

Our main result means, that for power-law MFD (contrary to 
exponential decay) we cannot put serious limits on the parameters
of decay with the ROSAT observations of INS candidates
as far as for all plausible models of power-law MFD
INSs from low velocity tail
are able to become accretors. For more detailed conclusions a NS census
for power-law MFD is necessary, similar to non-decaying and exponential
cases (Popov et al. 2000).

An interesting possibility of observing evolved accreting magnetars appear both
for the case of MFD and for constant field evolution. These
sources should be different from typical present day INS candidates observed
by ROSAT. Existence or absence of old accreting magnetars is  very
important for the whole NS astrophysics.

\vskip 0.2cm

\noindent
{\bf Acknowledgments}

\vskip 0.1cm

\noindent
We thank drs. Monica Colpi, Vasilii Gvaramadze and Roberto Turolla for
discussions. S.P. thanks University of Milan, University of Padova and
Astronomical Observatory of Brera for their hospitality.
This work was supported by the RFRB, INTAS and NTP ``Astronomy'' grants.



\begin{thebibliography}{}
\bibitem{}Alpar M.A. 1999, astro-ph/9912228
\bibitem{}Colpi M., Geppert U. and Page D. 2000, ApJL in press (astro-ph/9912066)
\bibitem{}Duncan R.C. and Thompson C. 1992, ApJ 392, L9
\bibitem{}Geppert U., Page D., Colpi M. and Zannias T. 2000, to be published
in the proceedings of the IAU Coll. 177 (astro-ph/9910563)
\bibitem{}Goldreich P. and Reisenegger A. 1992, ApJ 395, 250
\bibitem{}Haberl F., Pietsch W. and Motch C. 2000, A\&A in press (astro-ph/9911159)
\bibitem{}Konar S. and Bhattacharya D. 2000, to be published in the
proceedings of NATO/ASI on
'The Neutron Star - Black Hole Connection' (astro-ph/9911239)
\bibitem{}Konenkov D.Yu. and Popov S.B. 1997, PAZH 23, 569 (astro-ph/9707318)
\bibitem{}Kouveliotou C. et al. 1999, ApJ 510, L115 (astro-ph/9809140) 
\bibitem{}Lipunov, V.M. 1992, ``Astrophysics of Neutron Stars'' (NY: Springer \& Verlag)
\bibitem{}Lipunov V.M., Popov S.B. 1995, AZh 72, 711 (astro-ph/9609185)
\bibitem{}Marsden D., Lingenfelter R.E., Rothschild R.E. and Higdon J.C. 2000,
ApJ submitted (astro-ph/9912207) 
\bibitem{}Mereghetti S. and Stella L. 1995, ApJ 442, L17
\bibitem{}Mereghetti S. 1999, astro-ph/9911252
\bibitem{}Ne\"uhauser R., and Tr\"umper J.E. 1999, A\&A, 343, 151
\bibitem{}Perna R., Hernquist L. and Narayan R. 2000, ApJ submitted
(astro-ph/9912297)
\bibitem{}Popov S.B., Colpi M., Treves A., Turolla R., Lipunov V.M. and
Prokhorov M.E. 2000, ApJ 530, in press (astro-ph/9910114)
\bibitem{}Popov S.B. and Prokhorov M.E., 2000, A\&A in press (astro-ph/9908282)
\bibitem{}Popov S.E. and Konenkov D.Yu., 1998, Radiofizika 41, 28
(astro-ph/9812482)
\bibitem{}Schwope A.D., Hasinger S.G., Schwarz R., Haberl R. and Schmidt M.
1999, A\&A 341, L51 (astro-ph/9811326)
\bibitem{}Treves A., Turolla R., Zane S. and Colpi M. 2000, 
PASP in press (astro-ph/9911430)
\bibitem{}Urpin V.A., Geppert U. and Konenkov D.Yu. 1996, A\&A 307, 807
\end{thebibliography}
\end{document}